\documentclass[a4paper]{article}

\usepackage{INTERSPEECH2018}
\usepackage{balance}
\usepackage[colorinlistoftodos]{todonotes}

\title{A Survey Investigating Usage of Virtual Personal Assistants}
\name{Mateusz Dubiel$^1$, Martin Halvey$^1$, Leif Azzopardi$^1$}

\address{
  $^1$Dept. Computer and Information Sciences, The University of Strathclyde,  UK}
\email{mateusz.dubiel@strath.ac.uk, martin.halvey@strath.ac.uk, leif.azzopardi@strath.ac.uk}

\begin{document}

\maketitle
\begin{abstract}
Despite significant improvements in automatic speech recognition and spoken language understanding - human interaction with Virtual Personal Assistants (VPAs) through speech remains irregular and sporadic. According to recent studies, currently the usage of VPAs is constrained to basic tasks such as checking facts, playing music, and obtaining weather updates. In this paper, we present results of a survey (N = 118) that analyses usage of VPAs by frequent and infrequent users. We investigate how usage experience, performance expectations, and privacy concerns differ between these two groups. The results indicate that, compared with infrequent users, frequent users of VPAs are more satisfied with their assistants, more eager to use them in a variety of settings, yet equally concerned about their privacy.

\end{abstract}
\noindent\textbf{Index Terms}:  human-computer interaction, survey

\section{Introduction}\label{intro}

Although the capability to instruct devices to perform tasks via voice commands has been available since the 1950s \cite{davis1952automatic}, it is only in the last few years that the proliferation of smartphones made voice interfaces accessible to users worldwide. In comparison with earlier implementations, nowadays, voice technology is more robust to noise \cite{maganti2014auditory} and dialect variations \cite{biadsy2011automatic}, while voice search is reportedly three times faster than typing on a mobile device \cite{ruan2016speech}. Currently, virtual personal assistants (VPAs) are freely available on mobile devices (e.g. Microsoft Cortana, Google Assistant, Apple Siri), home appliances (e.g. Amazon Echo, Google Home), and as part of car systems (e.g. Nuance-developed Systems). According to recent surveys \cite{mikejeffs2018,rebeccasentance2016}, the availability of virtual assistants and consequently their aggregate usage has been increasing. However, despite their widespread availability, robustness and speed, the number of people who use VPAs on a regular basis remains relatively low. Another recent survey [5], found that 70\% of iPhone users consider their usage of Siri to be sporadic and limited to basic tasks.

In this paper, we present results of the survey on usage of VPAs. We focus on respondents’ experience with virtual assistants and compare differences in behaviour between frequent and infrequent users. There are three main factors of investigation, namely, (1) usage experience, (2) performance expectations, and (3) privacy concerns. The selection of the above three factors was motivated by a number of studies - carried out in the area of VPA evaluation, in which these factors were reported to determine frequency and intensity of VPA usage \cite{cowan2017can,easwara2015privacy,efthymiou2016evaluating,kiseleva2016understanding,luger2016like,moore2017spoken,moore2016progress,kiseleva2017evaluating}. Although the factors (1), (2) and (3) were previously considered in isolation, in our work, we analyse them together – looking at their overall impact on VPA usage patterns.

Our survey contributes to current research on virtual assistants by providing insights into interplay between user satisfaction and VPA usage frequency, and improves understanding of how different user groups currently use this technology. Based on the findings of recent research studies on evaluation of virtual personal assistants (discussed in Section \ref{related}), we formulated our research questions regarding current usage of VPAs. 

\begin{itemize}
\item \textbf{RQ1:} How do functionality, usage, and satisfaction differ between frequent and infrequent VPA users? 
\item \textbf{RQ2:} Do frequent and infrequent users vary in terms of their perception of current interaction capabilities and future performance expectations of their VPAs? 
\item \textbf{RQ3:} Are privacy concerns different between frequent and infrequent VPA users?
\end{itemize}

\section{Related Work}\label{related}

Intelligent Personal Assistants were originally developed to make the interaction with computer systems more human-like, enabling people to use natural language to manage their schedules, and access a variety of tasks and services \cite{van2007modelling}. The present generation of such assistants (e.g. Cortana, Siri, etc.), often referred to as Voice Powered Assistants (VPAs) is the focus of our study.
The growing popularity of VPAs and their improved accessibility can be attributed to the latest advances in speech technology. In recent years, the introduction of deep neural networks (DNN) for acoustic and language modelling has made ‘automatic speech recognition’ (ASR) systems more robust \cite{sercu2016very,shinohara2016adversarial}, while the implementation of the knowledge graph enhanced the ‘spoken language understanding’ (SLU) capabilities of question answering systems \cite{kumar2017knowledge}. 
Recent studies in the area addressed usage patterns of virtual assistants and problems with their adoption. The evaluation of VPAs focused on different aspects that affect their continued use, such as usage experience \cite{cowan2017can,kiseleva2016understanding,luger2016like}, performance expectations \cite{moore2016progress}, privacy concerns and social acceptability \cite{easwara2015privacy,efthymiou2016evaluating}. We discuss these studies and their findings below.\\

\noindent\textbf{Usage Experience}:
Kiseleva et al. \cite{kiseleva2016understanding} found that users tend to be more satisfied with using VPAs for simple tasks (such as device control) rather than more complex, multi-turn tasks (e.g. making travel arrangements etc.) where preserving the context is crucial. Despite being comprehensive, the study was limited to one type of virtual assistant (Cortana). Our survey adopts a broader perspective by probing users’ satisfaction with a variety of VPAs.
Cowan et al. \cite{cowan2017can}  focused on the experiences of infrequent VPA users and their reasons for not using VPAs on a regular basis. The feedback obtained from the focus groups indicates that privacy concerns over data usage, and lack of trust in the assistant's capability to perform the task are some of the main reasons for people not to use the technology on a regular basis.  Although Cowan et al.'s study was limited only to infrequent users of Siri, it can be argued that data permanency, ownership of data and limited human-like interaction abilities are the factors that are relevant across different devices. Thus, it would be worth extending the scope of investigation to other assistants. Our study applies a broad approach towards evaluation of VPAs by considering the feedback of different user groups (defined in Section \ref{background}). \\

\noindent\textbf{Performance Expectations:}
Luger and Sellen \cite{luger2016like} conducted a series of semi-structured interviews (14 participants) to get insights into users' expectations regarding the performance of voice-controlled assistants.  It was reported that users' poor understanding of how their VPA worked and lack of device feedback led to frustration and discontinued use. Our survey incorporates feedback from a larger (N = 118) and more diverse sample.

In a similar study, conducted by Sorensen \cite{sorensen2017expectations}, expectations of novice users were compared between two different chatbot systems (i.e. a human-like and a robot-like system). The results indicated that the system that asked questions, provided feedback and informed user about its current status was perceived as better meeting users’ performance expectations. While Sorensen's study dealt exclusively with text as the input method, in our work we focus on voice interaction.

Moore \textit{et al.} \cite{moore2016progress} analysed the opinions of members of the public expressed in two surveys on spoken language technology. One that compared the opinions of experts and non-experts, and another that evaluated the degree of usage of voice technology on a daily basis. The overall results suggest that the ordinary people (non-experts) are more optimistic about the future capabilities of voice technology. However, poor system accuracy and inadequate language understanding skills prevent regular usage. In current work, we ask our respondents about their perception of their VPA's current language capabilities, and expectations regarding future natural language and conversational capabilities. \\

\noindent\textbf{Privacy Concerns and Social Acceptability:}
Easwara and Vu \cite{easwara2015privacy} ran an online survey study which explored the impact of privacy and social acceptability concerns on the usage of VPAs. The results of the study indicated that people were more likely to interact with their devices via voice in private locations and when surrounded by the people who they knew. The two main reasons that participants did not use VPAs were ‘privacy concerns’ and ‘concerns over being judged by other people’. Another study examining privacy and social acceptability was conducted by Efthymiou and Halvey \cite{efthymiou2016evaluating} in the context of voice based smartwatch search. The findings indicated that voice search has low social acceptability when carried out in front of strangers, mostly due to privacy concerns. Our work builds up on previous studies \cite{easwara2015privacy,efthymiou2016evaluating} by incorporating questions on respondents' concerns over privacy and social acceptability of VPAs when used in private and public spaces.

In this paper, we replicate, collate and extend the prior work on VPAs to get better understanding of current usage and factors that drive VPA adoption. Our focus is to determine how voice interaction and VPA usage vary between frequent and infrequent user groups. Unlike previous research on virtual assistants, we do not limit our investigation to any particular device.

\section{Survey} Our online survey\footnote{https://tinyurl.com/voice-technology - Last Accessed on 23/02/2018} consisted of a mixture of open and closed questions. The welcome page of the survey informed respondents about the purpose of the study, the estimated completion time, and explained the voluntary nature of the survey. Before commencing, the respondents were asked to provide consent for their replies to be stored and analysed. In total, there were 29 questions for respondents who reported to have used a virtual assistant at least once in the past (VPA users) and 12 questions for those who did not (VPA non-users). In this paper, we only report the results for VPA users. 
The survey responses were collected between June 2017 and November 2017. 

\subsection{Background Information}\label{background}
At the beginning, the respondents were asked when they started to use their VPA and how frequently they used it for. The goal of these questions was to provide background information regarding device usage and to distinguish between different groups based on frequency of VPA usage. The initial analysis of data revealed 2 major groups i.e. 'frequent users' and 'infrequent users'. For the comparative purposes, based on usage behaviour, we defined respondents who reported to use VPA at least once a week as 'frequent users' (FU) and respondents who reported to use their VPA less than once a week as 'infrequent users' (IU). We are aware that this cut-off point is somewhat arbitrary, however, we chose it to provide enough data to facilitate comparison between these two user groups.

The rest of the survey was divided into 3 main parts, i.e. 'usage experience', 'performance expectations', and 'privacy concerns and social acceptability'.\\

\noindent\textbf{Usage Experience:}
The questions in the ‘device usability’ section seek to address our RQ1 - which explores the most frequently used VPA functionalities and users' satisfaction with them. The questions asked in this section, expand on the questions from the survey on the awareness of speech technology, reported in \cite{moore2017spoken}. Respondents are asked to comment on their last use of different VPA functionalities (measured on a Likert scale where '1' signifies 'never', '2' = 'more than a month ago', '3' = 'more than a week ago', '4' = 'within last week', and '5' means 'today'), and their satisfaction with each of the functionalities (measured on a Likert scale where '1' signifies 'very dissatisfied', 3 = 'neutral' and '5' means 'very satisfied'). The choice of basic functionalities used in the survey was motivated by the list provided on Google Blog \cite{googlesupport2018}. The complex functionalities were formulated based on tasks used in previous studies on VPAs \cite{kiseleva2016understanding} and \cite{moore2016progress}.\\

\noindent\textbf{Performance Expectations:}
This section of the survey seeks to address RQ2 (perception of current VPA's capabilities and expectations regarding future performance). There are two questions in this section: one that checks respondents’ expectations of VPA natural language skills (measured on a five point Likert scale where '1' stands for 'strongly disagree','3' = ‘neither agree nor disagree', and '5' is 'strongly agree'), and another that asks about the current perception of VPA conversation abilities (the same level of measurement). Both questions were motivated by the findings of Moore et al. \cite{moore2017spoken}, who posit that VPA usage will remain marginal unless assistants are equipped with human-like conversational skills. In our study we compare opinions of frequent and infrequent users to establish if there is a link between perceptions on VPAs performance and frequency of its usage.  \\

\noindent\textbf{Privacy Concerns and Social Acceptability:} The section seeks to address RQ3 (differences in privacy and social concerns between frequent and infrequent VPA users). First, respondents are asked to indicate their likelihood of VPA usage in different locations and in front of different audiences, (both measured on a five-point Likert scale where '1' signifies 'very unlikely', '3' = 'neutral', and '5' means ‘very likely’). Then, we ask if about respondents' privacy concerns (measured on Five-point Likert scale where '1' = 'not concerned at all' and '5' = 'extremely concerned'). The scales, types of audiences, and ranges of locations used in this part were adapted from \cite{easwara2015privacy}.

\subsection{Participants}
In total 215 respondents took part in the survey. After excluding incomplete replies, the data of 178 respondents was used in the analysis. Respondents were recruited via social networks (\textit{Twitter}, \textit{Facebook}, and \textit{LinkedIn}) and through information notes posted on notice boards on campuses of four major universities. In this paper we focus on VPA users i.e. people who used a voice activated personal assistant at least once. The VPA users group accounted of two-thirds of our sample (66\%, 118/178). In this group there were 58 males, 58 females and 2 respondents who classified their gender as 'other'. Younger respondents dominated the group - with under 35-year-olds accounting for over 74\% of its members (N = 88/118). All of the VPA users reported to be educated to a college level or higher. The survey average completion time was 15 minutes.

The majority of respondents (76\%, 90/118) have been using their VPA for at least 6 months. In terms of the frequency of usage - we divided respondents into ‘frequent users', i.e. the respondents who use their VPA at least once a week (45\%, 53/118) and ‘infrequent users’, the ones who use their VPA less than once a week (55\%, 65/118). Among the VPA users, 32\% used Apple Siri, 31\% used Google Assistant, 18\% used Amazon Echo and 19\% used other device. Most of the respondents used their VPA in English (~86\%, 101/118), followed by Italian (6\%, 7/118) and Polish (2.5\%, 3/118). In terms of first language, there were 20 different language groups. The biggest groups were: English (25\%, 47/188),  Polish (12\%, 15/118), Italian (9\%, 11/118), Spanish (9\%, 9/118) and Greek (8\%,8/118). It should be noted that although our respondents spoke various languages, given that our goal is to provide a broad overview of VPA usage, no attempts were made to balance this factor.   

\begin{figure}[h!]
  \centering
  \includegraphics[width=\linewidth]{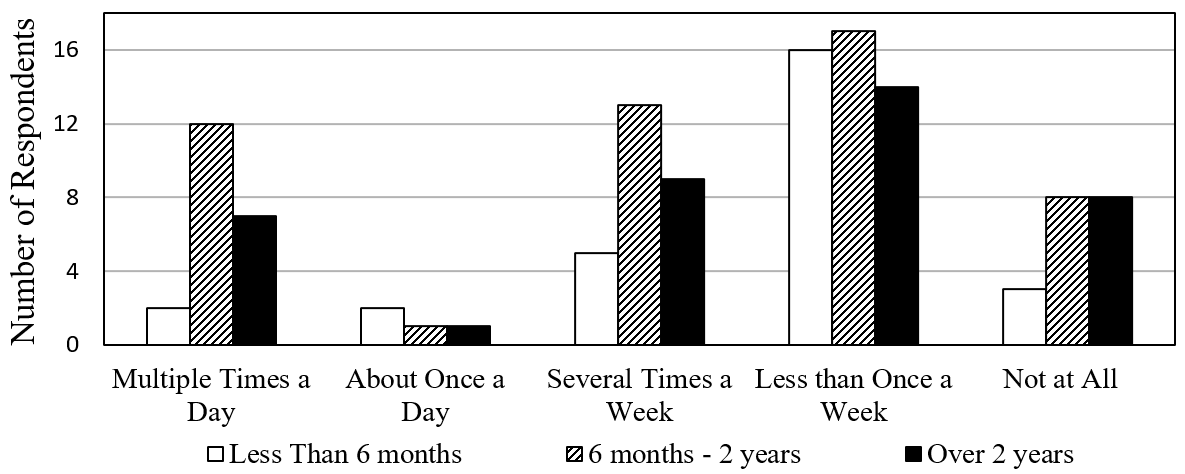}
  \caption{Frequency of VPA usage based on duration of usage.}
  \label{freq}
\end{figure}

A detailed breakdown of data indicates that users who have been using their VPA for longer are also more likely to do so more frequently (see Figure \ref{freq} for details). The proportion of respondents using their assistant multiple times every day is considerably higher among the users who have used their VPA for six months. However, for the most experienced users (the ones who have been using VPA for more than two years) the frequency of usage goes down slightly, as compared to those who have been using their VPA between 6 months and 2 years.

\section{Results}
In this section we report results of our survey on VPAs. When reporting results, unless stated otherwise, we use \textit{Mann Whitney U Test} \cite{nachar2008mann} for pairwise comparisons because the majority of our data is not normally distributed. \\

\begin{figure*}[h!]
\centering
  \includegraphics[width=.95\linewidth]{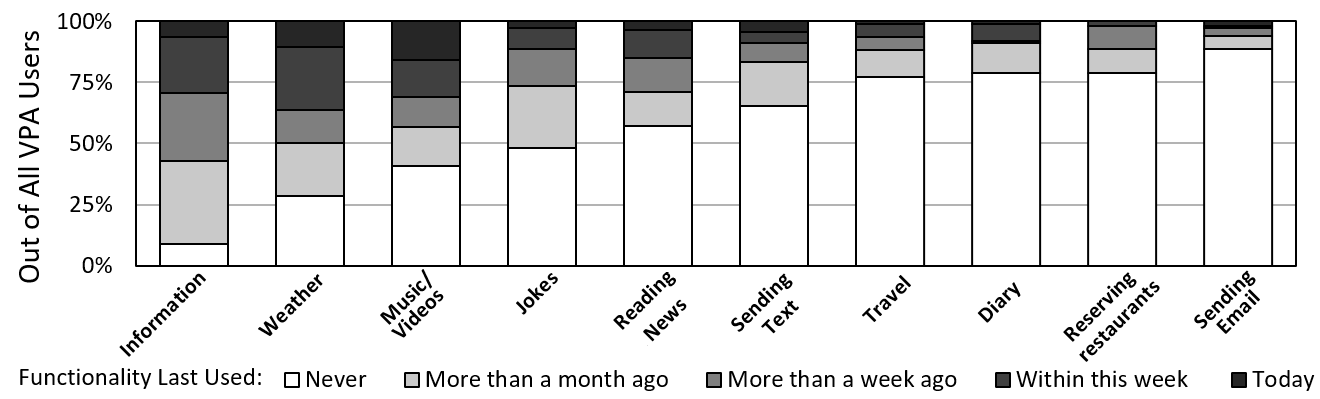}
  \caption{: Frequency of usage of VPA functionalities. The functionalities are presented in a descending order from the ‘most frequently used’ to the ‘least frequently used’.}
  \label{frequency}
\end{figure*}

\noindent\textbf{Usage Experience:} A full breakdown of usage frequency of VPA functionalities is presented in Figure \ref{frequency}. The results of a cross-comparison indicate that, overall, members of the ‘frequent group’ (respondents who used VPA at least once a week) were more satisfied with their VPA (p = 0.001), its ‘videos/music playing’ (p = 0.011), and ‘weather checking’ (p = 0.05) functions  than ‘infrequent users’ (respondents who used VPA less than once a week). More details are provided in Table \ref{tab2}. 

\begin{table}[h!]
\centering
\caption{Comparison of VPA Functionality Satisfaction Measured on a Five Point Likert Scale (where ‘1’ signifies ‘very dissatisfied’, ‘3’ = ‘neutral, and ‘5’ signifies ‘very satisfied’). Note: FU = ‘Frequent Users’, IU = ‘Infrequent Users’
 * indicates p = 0.05, ** indicates p $<$ 0.001}
\label{tab2}
\begin{tabular}{lll}
\hline
\multicolumn{1}{c}{\multirow{2}{*}{\textbf{Functionality}}} & FU (N = 53) & IU (N = 65)   \\ 
\multicolumn{1}{c}{}              & Med/M/SD    & Med/M/SD      \\ \hline
Looking for Information           & 4/3.37/1.04 & 4/3.35/1.16   \\
Managing Diary                    & 3/3.06/1.05 & 3/2.7/1.16    \\
Travel Arrangements                 & 3/3.24/1.04 & 3/3/1.02      \\
Reserving Restaurants             & 3/3.24/1.04 & 3/2.92/1.08   \\
Controlling Devices               & 4/3.94/1.12 & 3/2.83/1.16   \\
Keeping up to Date                & 4/3.42/1.16 & 3.5/3.23/1.17 \\
\textbf{Checking Weather*}                 & \textbf{4/4.22/.88}  & \textbf{4/3.76/1.14}   \\
Sending an Email                  & 3/2.88/1.13 & 2.5/2.4/1.35  \\
Sending a Text Message            & 4/3.33/1.11 & 3/3.14/1.95   \\
\textbf{Playing Music/Videos*}             & \textbf{4/3.98/1.1}  & \textbf{3/3.2/1.24}    \\
Telling Jokes                     & 4/3.48/.962 & 3/3.12/1.17   \\
\textbf{Overall** }                        & \textbf{4/3.71/.81}  & \textbf{3/3.04/1.05}   \\ \hline
\end{tabular}
\end{table}

\begin{table}[h!]
\centering
\caption{VPA usage likelihood Measured on a Five Point Likert Scale (where ‘1’ signifies ‘very unlikely’, ‘3’ = ‘neither likely nor unlikely', and ‘5’ signifies ‘very likely’. Note:'**' indicates p $<$ 0.001. Bonferroni adjusted alpha= 0.003 .}
\label{tab3}
\begin{tabular}{lll}
\hline
\multicolumn{3}{c}{\textbf{Usage Likelihood}}                                      \\ \hline
\multicolumn{1}{c}{\multirow{2}{*}{\textbf{Location}}} & FU (N = 53) & IU (N = 65) \\
\multicolumn{1}{c}{}                          & Med/M/SD    & Med/M/SD    \\ \hline
\textbf{Home**  }                                        & \textbf{5/4.45/1.1 } & \textbf{4/3.75/1.41} \\
Car (Driving)                                 & 4/2.96/2    & 3/2.34/2.1  \\
Pub                                           & 0/.76/.97   & 0/.65/1.05  \\
Pavement                                      & 2/1.82/1.6  & 1/1.35/1.59 \\
Public Transport                              & 1/1.04/1.26 & 0/.89/1.37  \\
Work                                          & 2/1.86/1.55 & 1/1.6/1.58  \\ \hline
\multicolumn{1}{c}{\multirow{2}{*}{\textbf{Context}}}  & FU (N = 53) & IU (N = 65) \\
\multicolumn{1}{c}{}                          & Med/M/SD    & Med/M/SD    \\ \hline
\textbf{Alone** }                                        & \textbf{5/4.71/.5}   & \textbf{4/4.15/.99}  \\
\textbf{Family**}                                       & \textbf{4/3.8/1.47}  & \textbf{4/2.94/1.5}  \\
Work Colleagues                               & 2/2.31/1.56 & 2/1.9/1.61  \\
\textbf{Partner**}                                       & \textbf{4/4.06/1.35} & \textbf{3/2.71/1.71} \\
Friends                                       & 4/3.67/1.37 & 4/3.06/1.53 \\
Strangers                                     & 1/1.59/1.4  & 1/1/1.73    \\ \hline
\end{tabular}
\end{table}

\noindent\textbf{Performance Expectations:}
No statistically significant differences in perception of current VPA capabilities, and VPA performance expectations were recorded between ‘frequent’ and ‘infrequent’ users. However, the frequent users were generally more in favour of VPA being able to recognise their interruptions (p = 0.54). The full results are presented in Table \ref{capa}.  
\begin{table}[h!]
\centering
\caption{VPA Current (Top) and Expected (Bottom) Language Capabilities. Statements are Measured on a Five Point Likert Scale (where 1 = 'Strongly Disagree' and 5 = 'Strongly Agree'). Note: FU = ‘Frequent Users’, IU = ‘Infrequent Users’.}
\label{capa}
\begin{tabular}{lll}
\hline
\multicolumn{1}{c}{\multirow{2}{*}{\textbf{Current Capabilities}}} & FU (N = 53)                  & IU (N = 65)                  \\ 
\multicolumn{1}{c}{}                 & Med/M/SD                    & Med/M/SD                  \\ \hline    
I need to speak differently          & 3/2.91/1                     & 3/2.91/1.1                   \\
VPA misunderstands me (NLU)                & 2/3/1.41                     & 2/3.09/1.3                   \\
VPA struggles with accent (ASR)                 & 2/3/1.41                     & 4/3.45/1.21                  \\
Results are irrelevant               & 4/3.36/1.36                  & 2/2.82/1.17                  \\ \hline
\multicolumn{1}{c}{\multirow{2}{*}{\textbf{Expected Capabilities}}}                     & FU (N = 53)                  & IU (N = 65)                  \\ 
                                     & \multicolumn{1}{l}{Med/M/SD} & \multicolumn{1}{l}{Med/M/SD} \\ \hline
Should Recognise Interruptions              & 5/4.45/.87                   & 4/4.09/0.83                   \\
Should be Human-Like                   & 5/4.3/1.2                    & 4/3.36/1.36                  \\
Should Have Personality                     & 4/4.09/1.3                   & 4/3.09/1.3                   \\
Should Ask More Questions                   & 4/4.09/0.83                   & 4/3.09/1.3                   \\ \hline
\vspace*{-4mm}
\end{tabular}
\end{table}

\noindent\textbf{Privacy Concerns and Social Acceptability:} Table \ref{tab3} presents VPA usage likelihood. Frequent users are significantly more likely to use their VPA at home, when alone, when with their family or with partner. As for privacy, both frequent and infrequent users express similar levels of concerns. For full breakdown of usage contexts and locations see Table \ref{tab5}.

\begin{table}[h!]
\centering
\caption{VPA - Privacy Concerns  measured on a 1 to 5 Likert scale where '1' signifies 'not concerned at all', and '5' signifies 'really concerned'. Note: FU = ‘Frequent Users’, IU = ‘Infrequent Users’. Bonferroni adjusted alpha=0.003}
\label{tab5}
\begin{tabular}{lll}
\hline
\multicolumn{3}{c}{\textbf{Privacy Concerns}}                                                                              \\ \hline
\multicolumn{1}{c}{\multirow{2}{*}{\textbf{Location}}} & FU (N = 53)                     & IU (N = 65)                     \\
\multicolumn{1}{c}{}                          & Med/M/SD                        & Med/M/SD                        \\ \hline
Home                                          & 2/2.17/1.32                     & 2/2.08/1.29                     \\
Car (Driving)                                 & 2/1.78/1.22                     & 2/2.05/1.26                     \\
Pub                                           & 4/3.27/1.45                     & 4/3.45/1.86                     \\
Pavement                                      & 3/3.1/1.78                      & 4/3.35/1.2                      \\
Public Transport                              & 4/3.4/1.41                      & 4/3.72/1.26                     \\
Work                                          & 3/3/1.3                         & 4/3.42/1.25                     \\ \hline
\multicolumn{1}{c}{\multirow{2}{*}{\textbf{Context}}}  & FU (N = 53)                     & IU (N = 65)                     \\
\multicolumn{1}{c}{}                          & Med/M/SD                        & Med/M/SD                        \\ \hline
Alone                                         & \multicolumn{1}{l}{1/1.88/1.26} & \multicolumn{1}{l}{1/1.98/1.27} \\
Family                                        & \multicolumn{1}{l}{2/2.35/1.28} & \multicolumn{1}{l}{2/2.48/1.15} \\
Work Colleagues                               & \multicolumn{1}{l}{3/2.85/1.24} & \multicolumn{1}{l}{3/3.14/1.2}  \\
Partner                                       & \multicolumn{1}{l}{2/2.4/1.36}  & \multicolumn{1}{l}{2/2.46/1.22} \\
Friends                                       & \multicolumn{1}{l}{2/2.6/1.28}  & \multicolumn{1}{l}{2/2.52/1.22} \\
Strangers                                     & \multicolumn{1}{l}{4/3.56/1.36} & \multicolumn{1}{l}{4/3.82/1.22} \\ \hline
\vspace*{-9mm}
\end{tabular}
\end{table}

\section{Conclusions and Future Work}
This paper has presented results of a survey on current usage of VPAs with focus on differences between frequent and infrequent users in terms of usage experience, performance expectations and privacy concerns. Based on the obtained results we address our research questions (formulated in Section \ref{intro}).

\textbf{RQ1}: As indicated by the quantitative data from questions on ‘usage experience’ - most of the respondents currently use their VPAs for simple tasks such as factoid queries, weather updates or playing music, while many functions are hardly used or completely unexplored (see Figure \ref{frequency} for details). We find that satisfaction varies between different users groups - with frequent users being significantly more satisfied with their VPAs overall (as presented in Table \ref{tab2}). This indicates a potential link between user satisfaction and VPA usage frequency.

\textbf{RQ2:} We observe that both frequent and infrequent VPA users have similar perceptions and expectations regarding performance of their virtual assistants. However, overall, the latter group tend to be more concerned with ASR rather than NLU capabilities of their VPAs (see Table \ref{capa} for details). This finding suggests that, despite recent technological developments, ASR is still perceived as a barrier to frequent usage of VPAs. 

\textbf{RQ3:} Both frequent and infrequent users seem to be equally concerned about their privacy when talking to VPAs in various settings. Yet, despite these concerns, frequent users are more eager to use their VPAs in front of different audiences including family members, and partners (as shown in Table \ref{frequency}). 

We find it particularly interesting that ASR is currently considered as one of the major concerns of infrequent users of VPAs. Thus, we intend to investigate this aspect further in our future work. In order to obtain more insights on perception of VPA speech recognition performance at individual user level, we will conduct diary studies of day-to-day VPA usage and follow them up with ethnomethodological analysis. 
 
In summary, we provided a broad overview of current VPA usage and highlighted some perceived barriers to regular use. The study opens up new research avenues that can be explored in order to get better understanding of human-VPA interaction.

\bibliographystyle{IEEEtran}
\balance
\bibliography{mybib}

\end{document}